\documentclass[prl,twocolumn,amsmath,amssymb,superscriptaddress,prl,floatfix]{revtex4}
\usepackage{color} 
\usepackage{graphicx}
\usepackage{dcolumn}
\usepackage{bm}
\usepackage{longtable}
\usepackage{textcomp}
\usepackage{epstopdf}
\usepackage{makeidx}     

\begin{document}

\setlength{\parskip}{0 pt}


\bibliographystyle{unsrt}

\title{Creation of X-Ray Transparency of Matter by Stimulated Elastic Forward Scattering}

\author{J. St\"ohr}
\email{stohr@slac.stanford.edu}
\affiliation{SLAC National Accelerator Laboratory, 2575 Sand Hill Road, Menlo Park, CA 94025, USA}

\author{A. Scherz}
\email{andreas.scherz@xfel.eu}
\affiliation{European XFEL GmbH, Albert-Einstein-Ring 19, 22761 Hamburg, Germany}

\begin{abstract}
X-ray absorption by matter has long been described by the famous Beer-Lambert law. Here we show how this fundamental law needs to be modified for high-intensity coherent x-ray pulses, now available at x-ray free electron lasers, due to the onset of stimulated elastic forward scattering. We present an analytical expression for the modified polarization-dependent Beer-Lambert law for the case of resonant core-to-valence electronic transitions and incident transform limited x-ray pulses. Upon transmission through a solid, the absorption and dichroic contrasts are found to vanish with increasing x-ray intensity, with the stimulation threshold lowered by orders of magnitude through a super-radiative coherent effect. Our results have broad implications for the study of matter with x-ray lasers.
\end{abstract}
\maketitle

Non-linear interactions of intense electromagnetic radiation with matter have long been utilized in the microwave and optical regions to control nuclear and valence electronic transitions and have enabled breakthroughs in many fields of science, such as medical imaging, telecommunication or the creation and manipulation of novel states of matter. The natural extension of these techniques into the x-ray region had to await the availability of sufficiently bright x-ray sources in the form of x-ray free electron lasers. Over the last few years, several experiments performed with rather uncontrolled x-ray pulses of high intensity, produced through the self amplification of spontaneous emission (SASE) process \cite{emma:10}, have revealed the presence of high intensity effects due to electronic stimulation \cite{stimulation} or multiple ionization \cite{ionization}.

Here we discuss how x-ray transmission through matter can be modified in a controlled way by stimulated scattering effects induced by transform limited x-ray pulses now available through self-seeding \cite{ratner:15}. In contrast to stimulated \emph{inelastic} scattering \cite{stimulation, mukamel, patterson:10}, which requires pulses with a broad bandwidth that covers the difference between excitation and de-excitation energies or multi-color pulses with separate ``pump'' and ``dump'' functions, we consider here the conceptually simpler case of \emph{elastic} stimulation which exists within the energy bandwidth of the incident beam itself. In this case stimulated x-ray scattering modifies the fundamental Beer-Lambert law  because of the direct link of x-ray absorption and resonant elastic scattering through the optical theorem. 

Of particular importance and interest are experiments that utilize \emph{resonant} electronic core-to-valence transitions since they exhibit large cross sections, and for solids provide elemental and chemical bonding specificity, and through their polarization dependence enable the determination of bond orientation \cite{stohr:NEXAFS} and the dichroic separation of charge and spin based phenomena \cite{stohr:Magnetism}. Resonant x-rays are widely utilized in x-ray absorption, x-ray scattering and coherent x-ray imaging experiments \cite{stohr:NEXAFS,stohr:Magnetism,eisebitt:04,ament:11}.

We derive the modified Beer-Lambert law  by utilizing the time-dependent density matrix approach where the evolution of the resonant core-valence two-level system is governed by the optical Bloch equations \cite{loudon}. An analytical solution is obtained for the case of incident transform limited x-ray pulses whose coherence time is much longer than the core hole lifetime. We apply our theory to the important case of 3d transition metal samples whose polarization dependent transmission exhibits both a charge and spin response, the latter through the x-ray magnetic circular dichroism (XMCD) effect. We find that for the prominent Co  L$_3$ absorption resonance at 778\,eV (wavelength of 1.6\,nm), stimulated decays begin to rob intensity from the dominant spontaneous Auger channel at an incident intensity of about 1\,mJ/cm$^2$/fs (1\,TW/cm$^2$), with the onset lowered by a super-radiative coherent scattering enhancement in the forward direction. At higher intensities the sample becomes increasingly transparent with the spin-based XMCD contrast disappearing sooner than the charge-based absorption contrast.

We follow the formalism of reference \cite{stohr:Magnetism} and, denoting the x-ray polarization by the labels $q\!=\!0$ for linear, $q\!=\!+$ for right and $q\!=\!-$ for left circular polarization, describe the polarization dependent x-ray response of a magnetic sample in terms of the atomic scattering length in the soft-x-ray approximation as $f^q(\vec{Q}\!=\!0)= r_0 Z \! + \! {f'}^q \! -  \mathrm{i} {f''}^q$, where $r_0$ is the Thomson scattering length and $Z$ the atomic number. The \emph{spontaneously} transmitted intensity through a sample of atomic number density $\rho_\mathrm{a}$ and thickness $d$ is given by the Beer-Lambert law
\begin{equation}
I^q_\mathrm{trans} =  I^q_0 \, \mathrm{e}^{ - 2 \lambda  {f''}^q \rho_\mathrm{a} d}
\label{Eq:Beer-Lambert-law}
\end{equation}
where  $I^q_\mathrm{trans}$ and $I^q_0$ are the polarization dependent transmitted and incident intensities and $\sigma^q_\mathrm{abs}  = 2 \lambda {f''}^q$ is the x-ray absorption cross section. Our x-ray scattering length formulation is related to the optical constants and the electric susceptibility through the complex refractive index $ \tilde{n}^q = 1-\delta^q + \mathrm{i} \beta^q \simeq 1 + \frac{1}{2}( {\chi'}^q + \mathrm{i} {\chi''}^q)$, where $\delta^q = \rho_\mathrm{a} \lambda^2 (r_0 Z + {f'}^q)/2 \pi$ and $\beta^q = \rho_\mathrm{a} \lambda^2 {f''}^q/2 \pi$. The resonant polarization dependent x-ray absorption cross section $\sigma^q_\mathrm{abs}  = 2 \lambda {f''}^q$ and the differential atomic elastic scattering cross section  $d \sigma^q_\mathrm{scat}/ d \Omega  = ({f'}^q)^2 + ({f''}^q)^2$  have a Lorentzian lineshape and are linked by the optical theorem which may be written as,
\begin{eqnarray}
{f''}^q  \! \!= \! \frac{\Gamma}{\Gamma^q_x } \frac{2 \pi}{\lambda }\!\left[({f'}^q)^2 \!\! + \!({f''}^q)^2 \right]  \! =  \!\frac{\Gamma_x^q}{\Gamma}\frac{\lambda}{2 \pi} \frac{ (\Gamma/2)^2} {(\hbar\omega \!- \! {\cal E}_0)^2 \!\! + \!(\Gamma/2)^2\!}
 \label{Eq:K-H-expressions-abs}
\end{eqnarray}
Here ${\cal E}_0$ is the resonant photon energy, $\Gamma= \Gamma^q_x +\Gamma_\mathrm{A}$ is the total spontaneous decay width, which in the soft x-ray region is dominated by the Auger width $\Gamma \simeq \Gamma_\mathrm{A}$  \cite{krause:79}. The polarization dependent radiative transition widths $\Gamma^q_x$ consist of a radial and angular part and can be calculated by \emph{ab initio} methods. We have derived their values for Fe, Co and Ni metal from experimental data, and they are listed in Table\,\ref{T:Dichroism-Fe-Co-Ni} .

\begin{table}[h]
\centering \caption[]{Polarization dependent parameters for the L$_3$ resonances of Fe, Co and Ni metals. Listed are the atomic number densities $\rho_\mathrm{a}$, the resonance energies and wavelengths, and the polarization dependent ($q \! = \!0,\pm$) peak experimental cross sections $\sigma^q_0$, assuming propagation along the magnetization direction. $ \Gamma^q_x$ is the polarization dependent dipole transition width which includes the number of valence holes $N_\mathrm{h}$, and $\Gamma$ is the natural decay energy width \cite{krause:79}. }
\renewcommand{\arraystretch}{1.1}
\setlength\tabcolsep{1.1pt}
\begin{tabular}{@{}lcccccccccccc@{}}
\hline\noalign{\smallskip}
 &    $\rho_\mathrm{a}$  & ${\cal E}_0$  &$\lambda_0$  &$\sigma^+_0$ &$\sigma^0_0$ &$\sigma^-_0$  & $ \Gamma^+_x $  & $ \Gamma^0_x $    & $ \Gamma^-_x $   & $\Gamma$    \\
        & $\left[\!\!\frac{ \mbox{\small atoms}}{ \mbox {nm}^3}\!\right]$   &[eV]    & [nm]   &[Mb]     &[Mb]     &  [Mb]    &[meV]              & [meV]             & [meV]      & [eV]            \\
\hline\noalign{\smallskip}
Fe  & 84.9  & 707  & 1.75  & 8.8    &  6.9   &  5.0    & 1.37    &  1.08     & 0.78     &  0.36     \\
Co  & 90.9  & 778  & 1.59  & 7.9    &  6.25  & 4.65    & 1.208   &  0.96     & 0.715    &  0.43     \\
Ni  & 91.4  & 853  & 1.45  & 5.1    &  4.4   & 3.7     & 0.675   &  0.575    & 0.48     &  0.48     \\
\hline
\end{tabular}
\begin{tabular}{@{}p{\textwidth}@{}}
\end{tabular}
\label{T:Dichroism-Fe-Co-Ni}
\end{table}

\begin{figure} [h]
\centering
\includegraphics[width=1.0\columnwidth]{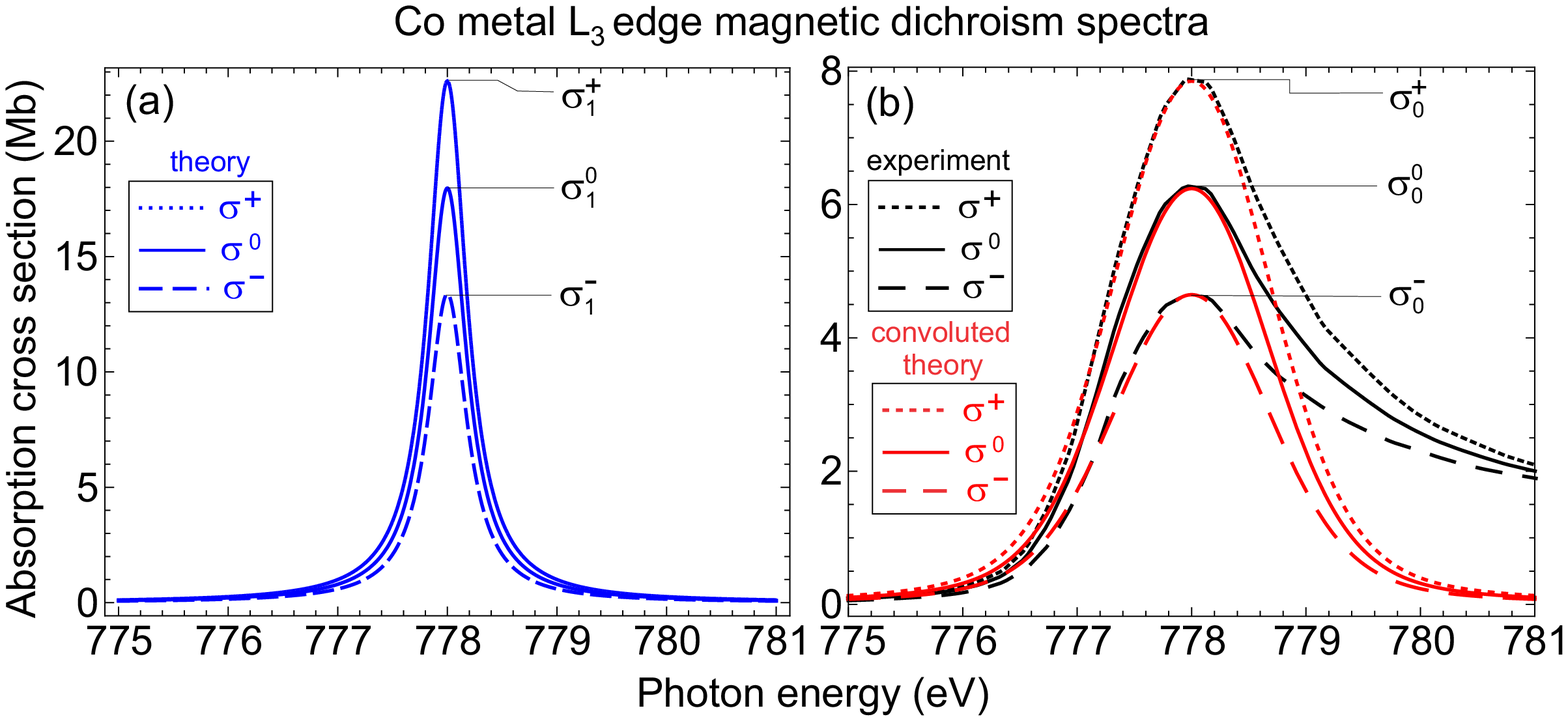}
\caption[]{(a) Polarization and photon energy dependent L$_3$ absorption cross sections $\sigma^q_\mathrm{abs}  = 2 \lambda {f''}^q$  for Co metal, calculated by use of Eq.\,\ref{Eq:K-H-expressions-abs} and the parameters $\Gamma$ and $\Gamma_x^q$ in Table\,\ref{T:Dichroism-Fe-Co-Ni}. (b) Comparison of the experimental dichroic cross sections (black lines) and the theoretical cross sections in (a) convoluted with a Gaussian of 1.4\,eV FWHM. Note that the corresponding blue and red curves have the same areas  but their peak values differ by a factor of  $\sigma^q_1/\sigma^q_0=2.9$.}
\label{Fig:Co-stim-dic-spectra}
\end{figure}

The polarization dependent Lorentzian x-ray absorption cross sections $\sigma^q_\mathrm{abs}  = 2 \lambda {f''}^q$  calculated with Eq.\,\ref{Eq:K-H-expressions-abs}  and the parameters for Co in  Table\,\ref{T:Dichroism-Fe-Co-Ni} are shown as blue curves in Fig.\,\ref{Fig:Co-stim-dic-spectra}\,(a). They were derived from fits of the experimental resonant cross sections by Voigt profiles as shown in Fig.\,\ref{Fig:Co-stim-dic-spectra}\,(b), consisting of a convolution of the natural Lorentzian lineshapes in (a) with a Gaussian of 1.4\,eV FWHM to account for the band-structure broadened $d$ valence states into which the $2p_{3/2}$ core electrons are excited.

For a sample of finite thickness $d$ and atomic number density $\rho_\mathrm{a}$, the transmitted intensity decays exponentially with the number of atoms in the beam $N_\mathrm{a}/A=\rho_\mathrm{a} d$ according to Eq.\,\ref{Eq:Beer-Lambert-law}. Since absorption and resonant scattering are related through Eq.\,\ref{Eq:K-H-expressions-abs}, the Beer-Lambert absorption law can also be derived by considering resonant elastic forward scattering. To do so one considers scattering by a thin atomic sheet so that the first Born approximation is valid. For a sheet thickness $\Delta \! \ll \! \lambda$ the spontaneously forward scattered fields are coherent and the transmitted field is given by,
\begin{eqnarray}
E^q_\mathrm{trans} = E^q_0 \, \mathrm{e}^{\mathrm{i}  k \Delta} \left\{1 -   \mathrm{i} \, \lambda \,\left [r_0 Z + {f'}^q - \mathrm{i} \, {f''}^q \right]  \, \rho_\mathrm{a} \Delta \right\}
\label{Eq:on-axis-transmitted-field}
\end{eqnarray}
Neglecting the non-resonant term $r_0 Z$, the intensity transmitted through the sample with $N_\mathrm{a}$ atoms in the beam of cross sectional area $A$ is,
\begin{eqnarray}
|E^q_\mathrm{trans}|^2\!\! = \! |E^q_0|^2 \!\left\{\! 1\! - \! 2 \lambda \frac{N_\mathrm{a}}{A} {f''}^q \!\! + \! \lambda^2 \frac{N^2_\mathrm{a}}{A^2} \! \left [({f'}^q)^2 \!\! + \! ({f''}^q)^2 \right] \! \right\}
\label{Eq:trans-int-scattering-picture}
\end{eqnarray}
The first term is the incident intensity and the second term is the absorption loss (minus sign) in linear response. Within the Born approximation, the absorption loss arises from the destructive interference of the incident field with the coherently forward scattered field. The third term is the forward scattered gain (plus sign) due to the coherent superposition of the fields scattered by the atoms in the sheet which scales with $N^2_\mathrm{a}$. In Eq.\,\ref{Eq:trans-int-scattering-picture} we have neglected the weak intensity $ (8\pi/3) N_\mathrm{a}[ ({f'}^q)^2 \! + \! ({f''}^q)^2 ]/A$ which is \emph{incoherently} scattered. The \emph{coherent} forward scattered intensity is larger by the enhancement factor,
\begin{eqnarray}
{\cal G}_\mathrm{coh}= \frac{3 }{8 \pi}N_\mathrm{a} \frac{\lambda^2}{A}
\label{Eq:enhancement-factor}
\end{eqnarray}
where $d \Omega_\mathrm{coh}=\lambda^2/A$ represents the solid angle of coherent forward scattering. The total field transmitted through a sample of arbitrary thickness $d = N \Delta$ is obtained by using the Darwin-Prins dynamical scattering formalism to sum Eq.\,\ref{Eq:on-axis-transmitted-field} over $N$ thin sheets, which yields the exponential Beer-Lambert law Eq.\,\ref{Eq:Beer-Lambert-law} \cite{henke:93}. Remarkably, for forward scattering, the longitudinal coherence length $\ell_\mathrm{c} \! \simeq \! \lambda^2/\Delta \lambda$ does not enter \cite{sinha:14}, and both Eq.\,\ref{Eq:Beer-Lambert-law} and ${\cal G}_\mathrm{coh}$ depend on the number of atoms $N_\mathrm{a}$ in the beam cross sectional area $A$, and not on the sample thickness $d$. ${\cal G}_\mathrm{coh}$ is thus the same for a thin film with $\rho_\mathrm{a} \simeq 100$\,atoms/nm$^3$ and typical thickness $d \! \simeq \!  1/(\sigma_\mathrm{abs} \rho_\mathrm{a}) \! \simeq \! 20$\,nm and a gas sample of the same atoms of density $\rho_\mathrm{a} \simeq 0.01$\,atoms/nm$^3$ and a much larger thickness of $d \simeq 200\,\mu$m.

As the incident intensity is increased, the Kramers-Heisenberg perturbation theory (see \cite{stohr:Magnetism}) leads to un-physical results since it does not account for population changes in the excited state. This is overcome by the density matrix formalism which directly calculates the time-dependent ground and excited state populations \cite{loudon}, which we shall denote $\rho_{11}(t)$ and $\rho_{22}(t)=1-\rho_{11}(t)$, respectively. The populations are obtained as the solutions of the optical Bloch equations.

In the presence of stimulation, we can write the atomic scattering length as the sum of a spontaneous (subscript ``0'') and stimulated non-linear (subscript ``NL'') part according to,
\begin{eqnarray}
{f'}^q  = r_0 Z + {f'}_0^q + {f'}_\mathrm{NL}^q,~~~{f''}^q = {f''}_0^q + {f''}_\mathrm{NL}^q
\label{Eq:f-NL-definition}
\end{eqnarray}
and Eq.\,\ref{Eq:Beer-Lambert-law} is replaced by,
\begin{equation}
I^q_\mathrm{trans} =  I^q_0 \,  \mathrm{e}^{ - 2 \lambda \left({f''_0}^q + 2 {f''}^q_{\!\!\!\mathrm{NL}} \right ) \rho_\mathrm{a} d}
\label{Eq:Stim-Beer-Lambert-law}
\end{equation}
The spontaneous absorption cross section $\sigma_\mathrm{abs}=2 \lambda {f''}^q$ with ${f''}^q ={f''_0}^q$ in Eq.\,\ref{Eq:Beer-Lambert-law} becomes the coherence time and intensity dependent expression,
\begin{eqnarray}
\sigma_\mathrm{abs} \!=\! 2 \lambda \left[{f''}^q_0 + 2{f''}^q_\mathrm{NL}\right] = 2 \lambda {f''}^q_0 \left[ 1 - 2 \, \rho^q_{22}(\tau_\mathrm{c})\right]~~
\label{Eq:effective-abs-cross-section}
\end{eqnarray}
Here ${f''}^q_0$ is given by the spontaneous expression Eq.\,\ref{Eq:K-H-expressions-abs}, and $\rho^q_{22}(\tau_\mathrm{c})$ is the excited state population obtained from the time dependent solution of the optical Bloch equations \cite{loudon}, by integration over the coherence time $\tau_\mathrm{c}$ of the incident x-rays.

\begin{figure} [h]
\centering
\includegraphics[width=1.0\columnwidth]{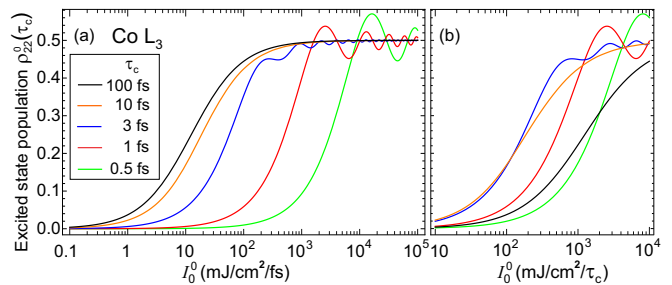}
\caption[]{(a) Excited state population $ \rho^0_{22}(\tau_\mathrm{c})$  as a function of the linearly polarized incident intensity $I^0_0$ for different coherence times $\tau_\mathrm{c}$ of the incident pulses for the L$_3$ edge of a 20\,nm thick Co metal film. We assumed resonance excitation, $\hbar\omega = {\cal E}_0$, linearly polarized light and experimental peak cross sections as discussed in the text. (b) $ \rho^0_{22}(\tau_\mathrm{c})$ as a function of the incident fluence per  coherence time of the pulse.}
\label{Fig:population-pulse-length-dependence}
\end{figure}

In general, the Bloch equations have to be solved numerically for $\rho^q_{22}(\tau_\mathrm{c})$. However, if the coherence time is much longer than the Auger decay time ($\hbar/\Gamma = 1.5$\,fs for Co $2p_{3/2}$), the excited state population reaches an equilibrium value (see Fig.\,\ref{Fig:population-pulse-length-dependence}) and the non-linear contribution is given by the analytical expression,
\begin{eqnarray}
{f''}^q_{\!\!\!\mathrm{NL}}  \!=\!  -  {f''_0}^q \! \underbrace{  \frac{ I^q_0  \Gamma^q_x \, {\cal G}_\mathrm{coh} \lambda^3  /(8  \pi^2 c )} {(\hbar\omega  \!- \! {\cal E}_0)^2 \! + (\Gamma/2)^2 \! +  I^q_0 \Gamma^q_x \, {\cal G}_\mathrm{coh} \lambda^3 /(4 \pi^2 c) } }_{\mbox{$\rho^q_{22}(\infty) $}}\,
 \label{Eq:fpp-beta-non-linear-q}
\end{eqnarray}
where $\rho^q_{22}(\infty) $ is the equilibrium excited state population in the limit $\tau_\mathrm{c}\rightarrow\infty$. The non-linear contribution ${f''}^q_\mathrm{NL}$ is seen to have the opposite sign of the spontaneous contribution ${f''_0}^q$. In the limit of high incident intensity we simply have $\rho^q_{22}(\infty) =0.5$ and $2 {f''}^q_{\!\!\!\mathrm{NL}} = - {f''_0}^q$ and the sample becomes transparent.

The increase in excited state population $ \rho^0_{22}(\tau_\mathrm{c})$  for Co L$_3$ excitation as a function of incident intensity and different coherence times of the incident pulse is shown in Fig.\,\ref{Fig:population-pulse-length-dependence}. It was calculated by numerical solution of the optical Bloch equations, assuming a 20\,nm thick Co metal film and $\hbar\omega = {\cal E}_0$, with the spontaneous peak cross section $\sigma^0_0$ (Fig.\,\ref{Fig:Co-stim-dic-spectra}\,(b)), corrected for the $\tau_\mathrm{c}$-dependent energy bandwidth $ \Delta{\cal E} =  2 \hbar\sqrt{\pi \ln 4}/\tau_\mathrm{c}$ of the incident pulse \cite{goodman-SO}. With increasing coherence time $\tau_\mathrm{c}$ relative to the core hole life time, the Rabi oscillations in the excited state population are suppressed. The behavior of $ \rho^0_{22}(\tau_\mathrm{c})$ for our chosen  $\tau_\mathrm{c}$ values as a function of the total intensity per coherent pulse, in units of [mJ/cm$^2$/$\tau_\mathrm{c}$] is shown in Fig.\,\ref{Fig:population-pulse-length-dependence}\,(b). The stimulated threshold is seen to be lowest for coherent pulses in the 3-10\,fs range.

\begin{figure} [*t]
\centering
\includegraphics[width=0.8\columnwidth]{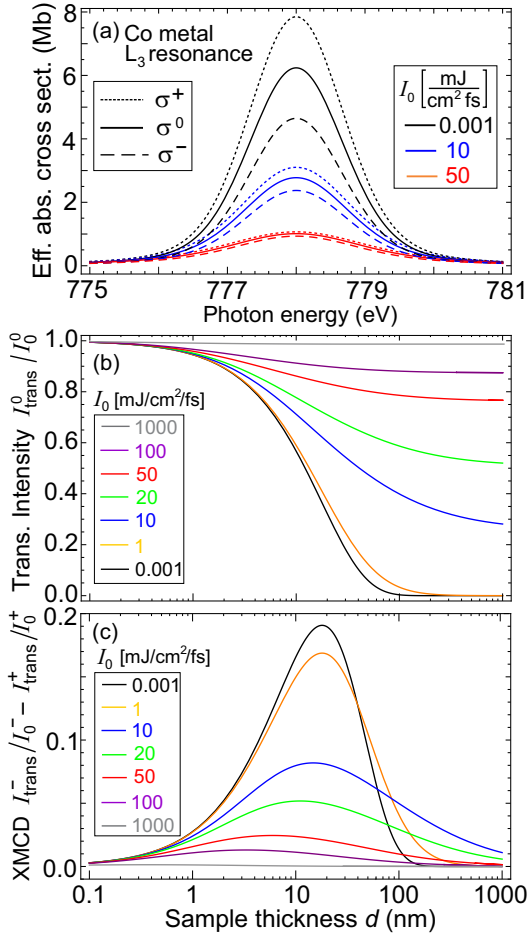}
\caption[]{(a) Change of the effective polarization dependent absorption cross section $2 \lambda \left({f''_0}^q + 2 {f''}^q_{\!\!\!\mathrm{NL}} \right )$ for the L$_3$ resonance in Co metal for three incident intensity values, assuming alignment of the x-ray propagation direction with the film magnetization and the long coherence time limit Eq.\,\ref{Eq:fpp-beta-non-linear-q}. The low intensity cross sections shown as black curves are nearly identical to the spontaneous ones shown in red in Fig.\,\ref{Fig:Co-stim-dic-spectra}. (b) Dependence of the transmission contrast as a function of sample thickness and incident intensity for resonant L$_3$ excitation of a  Co metal film due to charge absorption with linearly polarized light, according to Eq.\,\ref{Eq:Stim-Beer-Lambert-law}. (c) Same as (b) for the transmitted XMCD contrast. }
\label{Fig:Contrast-vs-thickness-intensity}
\end{figure}

Fig.\,\ref{Fig:Contrast-vs-thickness-intensity}\,(a) shows the effective polarization dependent absorption cross section for Co given by Eq.\,\ref{Eq:effective-abs-cross-section}  for three values of the incident intensity with ${f''}^q_{\!\!\!\mathrm{NL}}$ calculated according to Eq.\,\ref{Eq:fpp-beta-non-linear-q}. In  Fig.\,\ref{Fig:Contrast-vs-thickness-intensity}\,(b) we illustrate the thickness dependence of the absorption contrast (linear polarization) obtained from Eq.\,\ref{Eq:Stim-Beer-Lambert-law} for several values of the incident intensity. At low intensity the sample transmission decreases with increasing sample thickness $d$ due to absorption. However, with increasing intensity, the transmitted intensity at large $d$ is seen to decrease considerably slower due to stimulated forward scattering. The magnetic XMCD contrast, plotted in Fig.\,\ref{Fig:Contrast-vs-thickness-intensity}\,(c), first increases with thickness up to a maximum around $d \! = \! 1/(\sigma_\mathrm{abs} \rho_\mathrm{a}) \! = \!17$\,nm, corresponding to one x-ray absorption length, before it also decreases.

\begin{figure} [t]
\centering
\includegraphics[width=0.9\columnwidth]{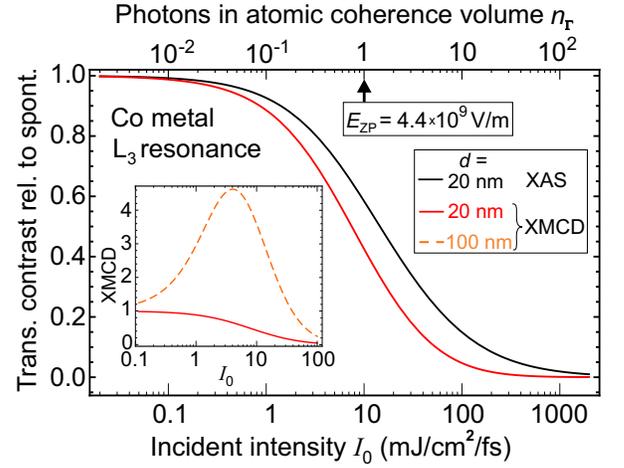}
\caption[]{ Dependence of the transmitted intensity according to Eq.\,\ref{Eq:Stim-Beer-Lambert-law} and Eq.\,\ref{Eq:fpp-beta-non-linear-q} in the presence versus absence of stimulation for a 20\,nm Co metal film as a function of incident intensity. The black curve represents the linear polarization or charge response $[I_\mathrm{trans}^0]_\mathrm{stim}/[I_\mathrm{trans}^0]_\mathrm{spon}$ and the red curve is the transmitted XMCD difference intensity $[I_\mathrm{trans}^- - I_\mathrm{trans}^+]_\mathrm{stim}/[I_\mathrm{trans}^- - I_\mathrm{trans}^+]_\mathrm{spon}$. The top scale is discussed in the text. The inset shows the relative transmitted XMCD contrast for film thicknesses of 20\,nm (red) and 100\,nm (dashed orange) as a function of incident intensity. }
\label{Fig:Abs-XMCD-decrease-with-intensity}
\end{figure}

Fig.\,\ref{Fig:Abs-XMCD-decrease-with-intensity} shows the dependence of the transmitted intensity for the stimulated relative to the spontaneous case as a function of the incident intensity, calculated for Co metal with $d=20$\,nm and assuming resonant excitation. Both the effective absorption cross section and the transmitted intensity reveal a strong dependence on the incident intensity, with the spin related XMCD contrast (red curve) vanishing faster than the charge related XAS contrast (black curve).

The inset reveals a particularly interesting thickness dependence of the transmitted XMCD intensity. For a thick sample of 100\,nm, the remaining small spontaneous XMCD contrast of about 1.5\%,  which according to Fig.\,\ref{Fig:Contrast-vs-thickness-intensity}\,(c) is greatly diminished by absorption, can actually be increased by nearly a factor of 5 upon stimulation.

The stimulated onset in Fig.\,\ref{Fig:Abs-XMCD-decrease-with-intensity} is predicted to be orders of magnitude lower than for the stimulated effects observed before \cite{stimulation} for SASE pulses with an average coherence time of about 0.5\,fs \cite{vartanyants:11}. This is due to our assumption of self-seeded pulses which besides coherence times of $\sim10$\,fs \cite{ratner:15} offer a jitter-free photon energy that can be resonantly tuned for maximum cross section. In addition, the elastic stimulation threshold is lowered by the coherent enhancement factor ${\cal G}_\mathrm{coh}$, which for L$_3$ excitation of a Co film is $\sim$\,500.

The dependence of the non-linear contribution on the incident intensity, given by Eq.\,\ref{Eq:fpp-beta-non-linear-q}, may also be expressed in terms of the number of incident photons contained in a specific volume. If the volume is chosen to be the coherence volume $V_{qk}$ per mode $qk$, then the associated number of photons $n_{qk}$ is referred to as the photon degeneracy parameter. The incident photons that stimulate electronic decays, however, need to be present during the total atomic clock decay time $\hbar/\Gamma$ which defines the sample-specific atomic decay volume $V_\Gamma$. The two coherence volumes are given by
\vspace*{-5pt}
\begin{eqnarray}
V_{qk} = \lambda^3 \frac{\hbar \omega}{ \Delta (\hbar \omega)},~~~ V_\Gamma = \lambda^3 \frac{ \hbar \omega}{2 \pi^2 \Gamma}
\end{eqnarray}
The number of stimulating photons $n_\Gamma$  in the volume $V_\Gamma$ is that in the well-known Kramers-Heisenberg stimulated correction term $1 + n_\Gamma$, and it can be expressed in terms of the incident polarization dependent intensity and field amplitude $E^q_0$ as,
\vspace*{-5pt}
\begin{eqnarray}
 n^q_\Gamma = \frac{1}{2 \pi^2 c} \frac{\lambda^3}{\Gamma} I^q_0=  \frac{\epsilon_0}{ \pi^2} \frac{\lambda^3}{\Gamma } |E^q_0|^2 =  \frac{|E^q_0|^2}{ |E_\mathrm{ZP}|^2}
\end{eqnarray}
On the right we have introduced the zero-point (ZP) field $E_\mathrm{ZP}$ responsible for spontaneous radiative decays. For $n^q_\Gamma=1$ the spontaneous and stimulated scattering intensities become the same, and the incident field $E^q_0$ is equally effective in driving decays as the ZP field $|E_\mathrm{ZP}|^2 = \pi^2\Gamma/(\epsilon_0\lambda^3)$ corresponding to one virtual photon in the volume $V_\Gamma$. This allows us to equate the intensity scale on the bottom of Fig.\,\ref{Fig:Abs-XMCD-decrease-with-intensity} with the number of photons $n_\Gamma$ on top of the figure, and for our case the ZP field has the value $E_\mathrm{ZP} \simeq 4.4\times 10^9$\,V/m.

Research at SLAC was supported through the Stanford Institute for Materials and Energy Sciences which is funded by the Office of Basic Energy Sciences of the U.S. Department of Energy. We would like to thank D. Higley, J. W. Goodman, and S. K. Sinha for clarifications of longitudinal coherence effects.

\end{document}